\documentclass[12pt,a4paper]{{amsart}}
\usepackage{verbatim} 
\usepackage{amssymb,amsfonts,latexsym,amscd}
\usepackage[dvips]{graphicx} 

\newcommand{\di}{\,\mathrm{d}}

\newcommand{\R}{\mathbb{R}}

\newtheorem{Lemma}{Lemma}[section]

\author{Helmut Linde} %
\title[Two-particle binding in a wave guide]{Geometrically induced two-particle binding in a wave guide}
\email{Helmut.Linde@gmx.de}
\address{Department of Physics, Pontificia Universidad Cat\'olica de Chile Casilla 306, Correo 22
Santiago, Chile.}
\thanks{This work was supported by CONICYT}

\begin{document}

\begin{abstract}
\noindent For mathematical models of quantum wave guides we show that in some situations two interacting particles can
be trapped more easily than a single particle. In particular, we give an example of a wave guide that can not bind a
single particle, but does have a geometrically induced bound state for two bosons that attract each other via a
harmonic potential. We also show that Neumann boundary conditions are `stickier' for two interacting bosons than for a
single one.
\end{abstract}

\maketitle
\section{Introduction}

Over the last two decades a considerable amount of research has been done on mathematical models for quantum wave
guides (see e.g. \cite{BGRS, DE, ELW, ESTV, EV96, EV97} and references therein). Typically a particle in such a
structure is modelled by a Schr\"odinger operator on some tube-like domain in two or three dimensions. The main object
of interest is the spectrum of these operators, and especially their low-lying eigenvalues which indicate the presence
of bound states for the particle. Such trapped modes have been proven to exist, e.g., for tubes with local
deformations, bends, or mixed boundary conditions. Much less is known though about the binding of several interacting
particles in such settings \cite{EV99, EZ, NTV95}. In \cite{EV99} Exner and Vugalter addressed the question how many
fermions can be bound in a curved wave guide if they are non-interacting or if they interact via a repulsive
electrostatic potential. It is clear that for these systems a smaller number of particles can be bound more easily than
a higher number of particles. In the present article we consider the somewhat opposite case and show that under certain
conditions two bosons with an attractive interaction can be bound more easily than one particle alone.

Our work is inspired by the analogous effect for Schr\"odinger operators\footnote{We choose units in which the Planck
constant $\hbar$ is equal to one.} in free space: Consider for a particle of mass $m$ the operator
$$H = -\frac{1}{2m} \Delta + V(x)$$
in $L^2(\mathbb R^n)$ with a non-trivial, compactly supported and bounded potential $V \le 0$. It is well known that
for $n > 2$ the attractive potential $V$ may be too weak to have bound states, i.e., $H$ may not have negative
eigenvalues. If this is the case, the same potential may still give rise to bound states of a system of two particles
that attract each other. This can be understood by physical intuition if one assumes that the two particles act in some
sense like one particle of the double mass. After all, as far as the existence of eigenvalues is concerned, doubling
the mass has the same effect as doubling the strength of the potential. In the present article we discuss whether an
analogous effect can occur for purely geometrically induced bound states in wave guides.

More precisely, we describe a quantum mechanical particle in a wave guide by the Dirichlet Laplacian $-\Delta$ in
$L^2(\Omega)$, where $\Omega$ is a straight strip or tube. The spectrum of this operator is purely continuous and
contains every real number above some threshold, which is the lowest eigenvalue of the Laplace operator on the cross
section of $\Omega$. It is known that geometrical perturbations like bending the tube or local deformations of the
boundary can give rise to eigenvalues of $-\Delta$ below this threshold. In analogy to the case of the Schr\"odinger
operator with a weak attractive potential, we ask the following question: Does a wave guide exist that doesn't have a
bound state for one particle, but that does have a bound state for a system of two interacting particles?

This question is not so easy to answer by physical intuition, because the existence or non-existence of geometrically
induced bound states for one particle doesn't depend on the mass of the particle in question. This means that the
intuitive `double mass argument' for two particles in an attractive potential doesn't apply to this situation. Despite
that, we will show in the following two sections that the answer to the question above is `yes' by giving an
appropriate example.

\section{Two-particle bound states in deformed wave guides} \label{SectionBulge}

We assume our wave guide to be the domain $\Omega \subset \R^2$ given by
$$\Omega = \{(x,y): |y| < f(x)\}$$
where
$$f(x) = \left\{\begin{array}{ll} \frac 12 & \textmd{for } |x| >  L/2,\cr%
\epsilon / 2 & \textmd{for } x = \pm \, L/2,\cr%
h / 2 & \textmd{for } |x| < L/2 \end{array} \right.$$%
with $L>0, h>1$ and $0<\epsilon <1$.
\begin{figure}
\includegraphics{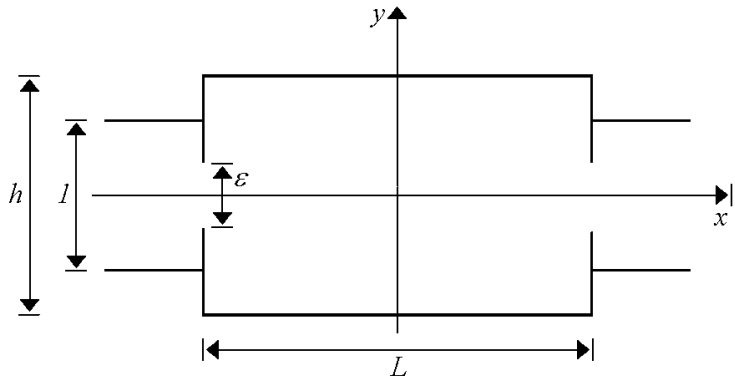}
\caption{Sketch of the wave guide $\Omega$}
\end{figure}
We impose Dirichlet conditions on $\partial \Omega$, which includes the `barriers' at $x= \pm L/2$. Our geometry can be
interpreted as a cavity of length $L$ and width $h$ coupled weakly (if $\epsilon$ is small) to two semi-infinite
straight wave guides. We choose to set $m = \frac 12$, such that the one-particle Hamiltonian is simply $H_1 =
-\Delta$. Then standard arguments imply that
$$\sigma_{\rm ess}(H_1) = [\pi^2,\infty).$$
Eigenvalues may occur depending on the choice of the parameters $L, h$ and $\epsilon$, but we will show:
\begin{Lemma} \label{Lemma1}
If $L^{-2} + h^{-2} > 1$ then for small enough $\epsilon$ there are no eigenvalues of $H_1$ below $\pi^2$, i.e., in
this case the wave guide has no one-particle bound states.
\end{Lemma}

On the other hand, we consider a system of two bosons of mass $m=\frac 12$, which interact via the harmonic potential
$$V = \alpha (x_1-x_2)^2 + \alpha(y_1-y_2)^2.$$
Here $x_i$ and $y_i$ are the particle coordinates and $\alpha>0$ is the interaction strength. To define the
self-adjoint Hamilton operator of the system we use the quadratic forms
\begin{eqnarray*}
h_{-\Delta}[\Psi] &=& \int_{\Omega\times\Omega} |\nabla\Psi|^2 \di x\di y \quad \textmd{and}\\
h_V[\Psi] &=& \int_{\Omega\times\Omega} V |\Psi|^2 \di x\di y,
\end{eqnarray*}
both defined on $C_0^\infty(\Omega\times\Omega)$. Then by \cite{D90}, Theorem 1.8.1, the sum of the two forms has a
closure $h_2$ with
$$h_2[\Psi] = \overline{h_{-\Delta}}[\Psi] + \overline{h_V}[\Psi]$$
for all $\Psi$ in
\begin{equation} \label{EqDomh2}
{\rm Dom}(h_2) = W_0^{1,2}(\Omega\times\Omega) \cap {\rm Dom}(\overline{h_V}).
\end{equation}
The positive self-adjoint operator associated with $h_2$ is
$$H_2 = -\partial_{x_1}^2 -\partial_{y_1}^2 -\partial_{x_2}^2 -\partial_{y_2}^2 + V$$%
in $L^2(\Omega\times\Omega)$.

\begin{Lemma} \label{Lemma2}
$\,$\\
\noindent a) For any choice of $L$ , $h$ and $\epsilon$ one has
$$\sigma_{\rm ess}(H_2) \subset  [\sqrt{2\alpha} + 2\pi^2, \infty).$$
\noindent b) There is a choice of the constants $L$ and $h$ with $L^{-2} + h^{-2} > 1$ such that
$$\inf \sigma(H_2) < \sqrt{2\alpha} + 2\pi^2$$
for every $\epsilon > 0$, i.e., the operator $H_2$ has a bound state.
\end{Lemma}

From the above lemmata we conclude that a wave guide exists that has no bound state for one particle, but does have a
geometrically induced bound state for two interacting particles.

A remark on the physical interpretation of this effect is in order. As mentioned above, the argument of two particles
acting like one of the double mass doesn't apply to geometrically induced bound states, since their existence is
mass-independent. To gain a physical intuition for our results anyway, we note that a bound state in a wave guide with
bulges can be seen as a trade-off between reduced kinetic energy in the transverse direction (due to the enlarged
cross-section) and increased kinetic energy in the longitudinal direction (due to the localization of the particle).
Consider now two particles that attract each other and that would in free space form a `molecule' with an average
distance $d$ between them. Assume for the case of our wave guide $\Omega$ that $d$ is considerably bigger than the
cavity width $h$, but considerably smaller than the cavity length $L$. This means that in their transverse movement the
two particles act rather as if they were independent of each other, thus receiving twice the energy decrease from the
enlarged cross-section. In longitudinal direction, on the other hand, the two particles in the cavity behave like one
particle of the double mass, such that the energy increase due to longitudinal localization is only half of what it
would be for one particle alone. It follows that the energy trade-off is more `favorable' for the system of two
interacting particles than for a single one.

\section{Two-particle bound states caused by Neumann boundary conditions}

If one introduces Neumann boundary conditions, an effect similar to the one described above happens even for particles
in only one dimension: Consider $H_3 = -\partial_x^2$ in $L^2(\R^+)$ with a Neumann condition at $x=0$. Then it is well
known that $\sigma_{ess}(H_3) = \R^+$ and $H_3$ has no eigenvalues. Nevertheless, the corresponding two-particle
Hamiltonian with an harmonic interaction turns out to have a bound state:

We define the potential $\hat V = \alpha |x_1-x_2|^2$ and the forms
\begin{eqnarray*}
\hat h_{-\Delta}[\Psi] &=& \int_{\R^+\times\R^+} |\nabla\Psi|^2 \di x_1 \di x_2 \quad \textmd{and}\\
\hat h_V[\Psi] &=& \int_{\R^+\times\R^+} \hat V |\Psi|^2 \di x_1 \di x_2,
\end{eqnarray*}
on the restrictions of the functions in $C_0^\infty(\R^2)$ to $\R^+\times \R^+$. Then we can take $h_4$ to be the
closure of $\hat h_{-\Delta} + \hat h_V$; and its associated self-adjoint operator is
$$H_4 = -\partial_{x_1}^2 - \partial_{x_2}^2 + \alpha |x_1-x_2|^2$$
on $L^2(\R^+ \times \R^+)$ with Neumann boundary conditions at $x_1=0$ and at $x_2=0$ (see, e.g., \cite{EE87}, page
340). The domain of $h_4$ is
\begin{equation}\label{EqDomh4}
{\rm Dom}(h_4) = W^{1,2}(\R^+\times\R^+) \cap {\rm Dom}(\overline{\hat h_V}).
\end{equation}
\begin{Lemma} \label{LemmaN1}
The operator $H_4$ has a bound state, i.e., an eigenvalue below the lower threshold of the essential spectrum.
\end{Lemma}

In view of Lemma \ref{LemmaN1} it is no surprise that wave guides exist which have no one-particle bound states, but
which do have a two-particle bound state induced by mixed boundary conditions. Omitting the proof, we only mention the
simple example of a straight tube with Dirichlet boundary conditions on the edge and an additional Neumann condition
imposed on one cross-section.

\section{Proofs of the results}

\begin{proof}[Proof of Lemma \ref{Lemma1}]
We introduce the operator $\tilde H_1$, which we define to be the Laplace operator on $\Omega$ with Dirichlet
conditions on $\partial \Omega$ and additional Neumann conditions on the set
$$\{(x,y): x=\pm L / 2 \textmd{ and } |y| < |\epsilon|\}.$$
To prove Lemma \ref{Lemma1} it is then sufficient to show that $\tilde H_1$ has no spectrum below $\pi^2$. With the
introduction of the new boundary conditions we have cut $\Omega$ into three separate domains: Two semi-strips
$\Omega^{+}$ and $\Omega^{-}$ in positive and negative $x$-direction, respectively, and the rectangle $\Omega^0 =
(-\frac L2,\frac L2)\times(-\frac h 2 ,\frac h 2)$. Thus $\tilde H_1$ is the orthogonal sum of the Laplace operators on
$\Omega^+$, $\Omega^-$ and $\Omega^0$ (subject to appropriate boundary conditions), and
$$\sigma(\tilde H_1) = \sigma(-\Delta_{\Omega^+}) \cup\sigma(-\Delta_{\Omega^-}) \cup \sigma(-\Delta_{\Omega^0}).$$%
One can convince oneself easily that
$$\sigma(-\Delta_{\Omega^+}) = \sigma(-\Delta_{\Omega^-}) = [\pi^2,\infty).$$
The spectrum of $-\Delta_{\Omega^0}$ is purely discrete and if we call $\lambda(\epsilon)$ its lowest eigenvalue then
$$\lambda(0) = \pi^2(h^{-2}+L^{-2}) > \pi^2.$$
We can now apply a theorem of Gadyl'shin \cite{G} to see that $\lambda(\epsilon) - \lambda(0)$ is of order
$\epsilon^2$, i.e., for small enough $\epsilon > 0$ we have $\inf \sigma(-\Delta_{\Omega^0}) > \pi^2$. Altogether this
means that
$$\inf \sigma(H_1) \ge \inf \sigma(\tilde H_1) = \pi^2 \quad \textmd{for small } \epsilon.$$
\end{proof}

\begin{proof}[Proof of Lemma \ref{Lemma2}, part a)]
Using the center of mass coordinates
\begin{equation}\label{Equw}
u = \frac 12 (x_2+x_1) \quad \textmd{and} \quad w = \frac 12(x_2-x_1)
\end{equation}
we rewrite $H_2$ in the form\footnote{In a slight abuse of notation we write $H_2$ for the two-particle Hamiltonian in
Euclidean coordinates and for its unitarily equivalent counterpart in center of mass coordinates.}
\begin{equation} \label{EqHc}
H_2 = -\frac 12 \partial_u^2 - \frac 12 \partial_w^2 + 4\alpha w^2 - \partial_{y_1}^2 - \partial_{y_2}^2 +
\alpha(y_2-y_1)^2.
\end{equation}
To estimate the spectrum of $H_2$ from below we introduce Neumann boundary conditions on
$$\{(u,w,y_1,y_2) : |w| = \beta\} \quad \textmd{and} \quad \{(u,w,y_1,y_2) : |w| < \beta , |u| = \beta+\frac L2\},$$
for some $\beta >0$, which turns $H_2$ into the orthogonal sum
$$\tilde H_2 = \left. H_2 \right|_{\{|w| > \beta\}} \oplus \left. H_2 \right|_{\{|w| < \beta, |u| < \beta + \frac L2\}}
\oplus \left. H_2 \right|_{\{|w| < \beta, |u| > \beta +\frac L2\}}.$$%

The spectrum of $\left. H_2 \right|_{\{|w| > \beta\}}$ can be estimated from below by $4\alpha \beta^2$ and the
spectrum of  $\left. H_2 \right|_{\{|w| < \beta, |u| < \beta + \frac L2\}}$ is discrete. By separation of variables the
spectrum of $\left. H_2 \right|_{\{|w| < \beta, |u| > \beta + \frac L2\}}$ is found to be purely continuous and its
lower threshold is equal to the lowest eigenvalue of the `transversal' operator
$$H_{t} = - \frac 12 \partial_w^2 + 4\alpha w^2 - \partial_{y_1}^2 - \partial_{y_2}^2 + \alpha(y_2-y_1)^2$$%
on $L^2((-\beta,\beta) \times (-1/2, 1/2)^2)$ with Neumann conditions at $|w| = \beta$ and Dirichlet conditions at
$|y_{1}| = 1/2$ and $|y_{2}| = 1/2$.  Neglecting the positive potential term $\alpha (y_2-y_1)^2$, we see that the
lowest eigenvalue of $H_t$ is bigger than $\lambda_\beta + 2\pi^2$, where $\lambda_\beta$ is the lowest eigenvalue of
the harmonic oscillator $- \frac 12 \partial_w^2 + 4\alpha w^2$ on $(-\beta,\beta)$ with Neumann boundary conditions.
Below we will show that for $\beta \rightarrow \infty$ the eigenvalue $\lambda_\beta$ converges to $\sqrt{2\alpha}$,
i.e., the lowest eigenvalue of the harmonic oscillator on $\R$. Consequently, for large enough $\beta$ the lowest
eigenvalue of $H_t$ is bigger than $\sqrt{2\alpha} + 2\pi^2$. Part a) of Lemma \ref{Lemma2} now follows from the fact
that $\tilde H_2 < H_2$ and the min-max principle.

It remains to show that $\lim_{\beta\rightarrow\infty} \lambda_\beta = \sqrt{2\alpha}$: Call $h_I = - \frac 12
\partial_w^2 + 4\alpha w^2$ the Hamiltonian of the harmonic oscillator on the interval $I \subset \R$ with
Neumann boundary conditions. Then
$$\lambda_\beta = \inf \sigma(h_{(-\beta,\beta)}) = \inf \sigma(h_{(-\infty,-\beta)} \oplus h_{(-\beta,\beta)} \oplus
h_{(\beta,\infty)}) \le \inf \sigma(h_{\R}) = \sqrt{2\alpha}.$$ %
The second step in the above chain of equalities follows from
$$\inf \sigma(h_{(-\beta,\beta)}) \le 4\alpha\beta^2 \quad \textmd{and}\quad \inf \sigma(h_{(-\infty,-\beta)}) = \inf \sigma(h_{(\beta,\infty)}) \ge 4\alpha\beta^2.$$

Next we show that $h_{(-\beta,\beta)}$ has a first eigenfunction that is symmetric, non-negative and decreasing in
$|w|$: Let $\phi_\beta$ be a normalized function such that $h_{(-\beta,\beta)}\phi_\beta=\lambda_\beta\phi_\beta$. We
may assume that $\phi_\beta$ is either symmetric or antisymmetric, since otherwise we can replace it by
$\phi_\beta(w)+\phi_\beta(-w)$. We write $\phi_\beta^\star$ for the symmetric decreasing rearrangement of $\phi_\beta$
(see \cite{T76} for the definition and properties of rearrangements). Then $\phi_\beta^\star$ is also normalized and
belongs to the form domain $W^{1,2}((-\beta,\beta))$ of $h_{(-\beta,\beta)}$. The min-max principle yields
\begin{eqnarray}\label{EqMist}
\lambda_\beta &\le& \int_{-\beta}^\beta \left(\frac 12 |{\phi_\beta^\star}'|^2 + 4\alpha w^2
{\phi_\beta^\star}^2\right) \di w\\
&\le& \int_{-\beta}^\beta \left(\frac 12 |{\phi_\beta}'|^2 + 4\alpha w^2 {\phi_\beta}^2\right) \di w =
\lambda_\beta.\nonumber
\end{eqnarray}
The second inequality in (\ref{EqMist}) follows from standard rearrangement theorems\footnote{The estimate
$\int_{-\beta}^\beta |{\phi_\beta^\star}'|^2  \di w\le \int_{-\beta}^\beta |{\phi_\beta}'|^2  \di w$ is a typical
rearrangement property. It is usually stated for functions that go to zero at the boundary of their domain, but it also
holds in the present case: Replacing $\phi_\beta(w)$ by $|\phi_\beta(w)|-|\phi_\beta(\beta)|$ and $\phi_\beta^\star(w)$
by $(|\phi_\beta(w)|-|\phi_\beta(\beta)|)^\star$ does not change the value of the integrals, and
$|\phi_\beta(w)|-|\phi_\beta(\beta)|$ is zero for $w = \pm \beta$ by (anti-) symmetry of $\phi_\beta$.}. The inequality
is strict (and thus a contradiction) unless $|\phi_\beta|$ is decreasing in $|w|$. This shows that $\phi_\beta$ can be
taken to be a non-negative symmetric eigenfunction to $\lambda_\beta$ that is decreasing in $|w|$. Then we have
$$\int_{-\beta}^\beta 4\alpha w^2 \phi_\beta^2(\beta) \di w \le \int_{-\beta}^\beta 4\alpha w^2 \phi_\beta^2(w) \di w \le
\lambda_\beta \le \sqrt{2\alpha}$$ %
and thus
\begin{equation} \label{EqVab}
\phi_\beta(\beta) \le 2^{-5/4} 3^{1/2} \alpha^{-1/4} \beta^{-3/2}.
\end{equation}
Now set
\begin{equation*}
\tilde\phi_\beta(w) = \left\{\begin{array}{ll} \phi_\beta(w) & \textmd{for } |w| \le \beta,\cr \phi_\beta(\beta)
(-|w|+\beta+1) & \textmd{for } \beta < |w| \le \beta +1,\cr 0 & \textmd{for } \beta +1< |w| \end{array}\right.
\end{equation*}
Then $\tilde\phi_\beta$ is in the form domain of $h_{\R}$ and we have
\begin{eqnarray}
\sqrt{2\alpha} &=& \inf \sigma(h_{\R}) \le \frac{\int_{\R} \left(\frac 12 \tilde\phi'_\beta(w)^2 + 4 \alpha w^2
\tilde\phi_\beta^2(w) \right) \di w}{\int_{\R} \tilde\phi_\beta^2(w) \di w} \nonumber\\
&\le& \lambda_\beta + 2\int_\beta^{\beta+1} \left(\frac 12 \phi_\beta^2(\beta) + 4\alpha w^2 \phi_\beta^2(\beta)
\right) \di w \nonumber\\
&=& \lambda_\beta + \phi_\beta^2(\beta) + \frac 83 \alpha  \phi_\beta^2(\beta) (3\beta^2+3\beta+1) \label{EqBBB}
\end{eqnarray}
In the penultimate step we used that $\int_{\R} \tilde\phi_\beta^2(w) \di w \ge \int_{-\beta}^\beta \phi_\beta^2(w) \di
w = 1$ and the Ritz-Rayleigh characterization of $\lambda_\beta$. From (\ref{EqVab}) we conclude that (\ref{EqBBB})
converges to $\lambda_\beta$ as $\beta\rightarrow \infty$ and therefore $\lim_{\beta\rightarrow\infty} \lambda_\beta =
\sqrt{2\alpha}$.
\end{proof}

\begin{proof}[Proof of Lemma \ref{Lemma2}, part b)]
We choose to fix the relations
\begin{equation}
\alpha = L^{-2} \quad \textmd{and}\quad h^{-2}+L^{-2} =: M > 1 \label{EqAA}
\end{equation}
between the parameters that describe our wave guide. We define the domain $\tilde\Omega$ as the set of all
$(x_1,y_1,x_2,y_2)$ that satisfy the conditions
$$u \in \left(-\frac {3L}8 , \frac {3L}8 \right), \quad w \in \left(-\frac L8 , \frac L8 \right), \quad
y_1,y_2 \in \left(-\frac h2, \frac h2\right),$$%
using the coordinates $u$ and $w$ as defined in (\ref{Equw}). One can check that $\tilde\Omega \subset
\Omega\times\Omega$. We now define the test function $\Psi$ by
$$\Psi = \left(\cos \frac {4\pi u}{3L}\right) \cdot (\phi(w)-C) \cdot
\left(\cos\frac{\pi y_1}{h}\right)\cdot \left(\cos\frac{\pi y_2}{h}\right)$$ %
on $\tilde \Omega$ and $\Psi = 0$ on $(\Omega\times\Omega)\backslash\tilde\Omega$, setting
$$\phi(w) = e^{-\sqrt{2\alpha} w^2} \quad \textmd{and} \quad C = \phi(L/8).$$
Because the function $\Psi$ is Lipschitz continuous, has a bounded support and vanishes at
$\partial(\Omega\times\Omega)$, we have $\Psi \in W_0^{1,2}(\Omega\times\Omega)$. Since the potential $V$, restricted
to the support of $\Psi$, is bounded, we also have $\Psi \in {\rm Dom}(\overline{h_V})$. By (\ref{EqDomh2}) this means
that $\Psi \in {\rm Dom}(h_2)$. In the center of mass coordinates the quadratic form of $H_2$ reads
\begin{eqnarray*}
h_2[\Psi] &=& \int \Bigl( \frac 12 (\partial_u\Psi)^2 + \frac 12 (\partial_w\Psi)^2 + (\partial_{y_1}\Psi)^2 +
(\partial_{y_2}\Psi)^2 \\
&& \quad \quad + 4\alpha w^2 |\Psi|^2 + \alpha (y_1-y_2)^2|\Psi|^2\Bigr) \di w\di u \di y_1 \di y_2.
\end{eqnarray*}
No we can apply the min-max principle with $\Psi$ as a test function to obtain
\begin{eqnarray}
\inf \sigma(H_2) \le \frac{h_2[\Psi]}{||\Psi||^2} &=& \frac{8\pi^2}{9L^2}  + \frac{2\pi^2}{h^2} +
\frac{\pi^2-6}{6\pi^2}\alpha h^2 \nonumber\\
&+& \frac{\int_{-L/8}^{L/8} (\frac 12 \phi'(w)^2 + 4\alpha w^2 (\phi(w)-C)^2) \di w}{\int_{-L/8}^{L/8} (\phi(w)-C)^2
\di w}.\label{EqA}
\end{eqnarray}
The last term can be estimated from above by
$$\frac{\int_{-L/8}^{L/8} (\frac 12 \phi'(w)^2 + 4\alpha w^2 \phi(w)^2 )
\di w}{\int_{-L/8}^{L/8} (\phi^2(w)-2C\phi(w)) \di w},$$ %
which can, after an integration by parts, be written as
\begin{eqnarray} && \frac{\sqrt{2\alpha} + \left(\int_{-L/8}^{L/8} \phi^2(w) \di w\right)^{-1}
[\frac 12 \phi(w)
\phi'(w)]_{-L/8}^{L/8}}{1 - 2 C \left(\int_{-L/8}^{L/8} \phi^2(w) \di w\right)^{-1} \int_{-L/8}^{L/8} \phi(w) \di w}\nonumber \\
&<& \frac{\sqrt{2\alpha}}{1 - 2 e^{-\sqrt 2 L/64} \left(\int_{-L/8}^{L/8} e^{-2\sqrt 2 L^{-1} w^2} \di w\right)^{-1}
\int_{-L/8}^{L/8} e^{-\sqrt 2 L^{-1} w^2} \di w}\nonumber
\end{eqnarray}
where in the last step we have used that $\alpha = L^{-2}$ and thus $C = e^{-\sqrt 2 L/64}$. Replacing $w$ by the new
variable $\tilde w = w/\sqrt L$ one can check that the product of the  two integrals in the last line converges to a
constant as $L \rightarrow \infty$. Therefore, the last term in (\ref{EqA}) can be estimated from above by
$\sqrt{2\alpha} + {\mathcal O}(L^{-1}e^{-\sqrt 2 L/64})$ for large enough $L$, which means that in view of (\ref{EqAA})
\begin{equation*}
\inf \sigma(H_2) < \sqrt{2\alpha}  + 2M\pi^2 - \frac{10 \pi^2}{9 L^2} + \frac{\pi^2-6}{6\pi^2 (L^2-1)} + {\mathcal
O}(L^{-1}e^{-\sqrt 2 L/64}).\label{EqB}
\end{equation*}
If we choose $L$ sufficiently large then the three last summands together are negative. If we then choose $M$
sufficiently close to one, we get independently of $\epsilon$
\begin{equation*}
\inf \sigma(H_2) < \sqrt{2\alpha}  + 2\pi^2,
\end{equation*}
proving part b) of Lemma \ref{Lemma2}.
\end{proof}

\begin{proof}[Proof of Lemma \ref{LemmaN1}]
In the center of mass coordinates $H_4$ acts in\goodbreak$L^2(\{(u,w):u>0,|w|<u\})$ and takes the form\footnote{Again
we abuse our notation and denote the two operators with respect to different coordinates by the same symbol $H_4$,
since they are unitarily equivalent.}
$$H_4 = -\frac 12 \partial_u^2 - \frac 12 \partial_w^2 + 4\alpha w^2.$$
Using a similar argument as in the proof of Lemma \ref{Lemma2}, part a), one can show that
$$\sigma_{ess}(H_4) = [\sqrt{2\alpha}, \infty).$$
It remains to prove that $H_4$ has an eigenvalue below $\sqrt{2\alpha}$. We call $\phi(w)$ the (positive and
normalized) lowest eigenfunction of the harmonic oscillator $-\frac 12
\partial_w^2+4\alpha w^2$ in $L^2(\R)$ and note that the corresponding eigenvalue is $\sqrt{2\alpha}$. We define the test function
\begin{equation*}
\Psi(u,w) =  \phi(w) e^{-\epsilon u} \quad \textmd{for } \,u>0, |w|<u\, \textmd{ and some } \epsilon > 0.
\end{equation*}
We have $\Psi \in W^{1,2}(\{(u,w):u>0,|w|<u\})$ and since $\Psi$ drops off exponentially for $u,|w|\rightarrow \infty$,
while $V$ is only quadratic, also $\Psi\in{\rm Dom}(\overline{\hat h_V})$ holds. Thus $\Psi$ is in the form domain
(\ref{EqDomh4}) of $H_4$ and we can apply the min-max principle \cite{RS4} to obtain
\begin{eqnarray*}
\inf \sigma(H_4) &\le& \frac{\int_{\,^{u>0}_{|w|<u}} \left( \frac 12 (\partial_u\Psi)^2 + \frac 12(\partial_w \Psi)^2 +
4\alpha
w^2\Psi^2\right) \di w \di u}{\int_{\,^{u>0}_{|w|<u}}  \Psi^2 \di w \di u}\\
&=& \frac 12 \epsilon^2 + \sqrt{2\alpha} + \frac{\int_{u>0} [\frac 12 \phi(w) \phi'(w)]_{-u}^u e^{-2\epsilon u} \di
u}{\int_{\,^{u>0}_{|w|<u}}  \Psi^2 \di w \di u}.
\end{eqnarray*}
In the last step we used an integration by parts in $w$ and the fact that $\phi$ satisfies the eigenvalue equation of
the harmonic oscillator. The last summand is negative since $\phi(w)$ is positive, symmetric and decreasing in $|w|$,
thus we have the estimate
\begin{eqnarray*}
\inf \sigma(H_4) &\le& \frac 12 \epsilon^2 + \sqrt{2\alpha} + \frac{\int_{u>0} [\frac 12 \phi(w) \phi'(w)]_{-u}^u e^{-2\epsilon u} \di u}{\int_{\,^{u>0}_{w\in \R}}  \Psi^2 \di w \di u}\\
&=& \frac 12 \epsilon^2 + \sqrt{2\alpha} + 2\epsilon \int_{u>0} \phi(u) \phi'(u) e^{-2\epsilon u} \di u\\
&=& \sqrt{2\alpha} + \epsilon \left( \frac 12 \epsilon + 2 \int_{u>0} \phi(u) \phi'(u) e^{-2\epsilon u} \di u\right)
\end{eqnarray*}
The integral in the last line is negative and its absolute value increases when $\epsilon$ goes to zero. Consequently,
for some small enough $\epsilon$ we have $\inf \sigma(\tilde H_4) < \sqrt{2\alpha}$, which proves Lemma \ref{LemmaN1}.
\end{proof}

\section*{Acknowledgments}

It is a pleasure for me to thank Rafael Benguria and Pavel Exner for their interest in this work and their helpful
comments. I am also very grateful to the referees for their valuable suggestions.

\end{document}